\documentclass[preprintnumbers,amsmath,amssymb,aps,prc,preprint,groupedaddress]{revtex4}

\usepackage{dcolumn}
\usepackage{bm}
\usepackage{epsfig}
\usepackage{newlfont}
\usepackage{citesort}

\usepackage{graphicx}

\begin{document}

\preprint{LA-UR-04-5803}

\begin{center}
\begin{large}
{\bf Electromagnetic excitation rates for nuclear isomers in a hot dense plasma}\\
\end{large}
\vspace{1 cm}
T.C. Luu, J.L. Friar, and  A.C. Hayes\\
{\it Theoretical Division, Los Alamos National Laboratory}\\
{\it MS-227 Los Alamos, NM 87545, USA}\\
\vspace{1 cm}
\end{center}
\begin{flushleft}
  Correspondence: Thomas Luu\\
  \hspace{3.05cm} Los Alamos National Laboratory\\
  \hspace{3.05cm} T-6 Division, MS B227\\
  \hspace{3.05cm} Los Alamos, NM  87545\\
  \hspace{3.05cm} Phone:  (505) 667-3612\\
  \hspace{3.05cm} Fax:    (505) 664-0007\\
  \hspace{3.05cm} tluu@lanl.gov\\
\end{flushleft}
\vspace{1 cm}
18 total pages\\
3 tables\\
1 figure

\newpage

\title
{\bf Electromagnetic excitation rates for nuclear isomers in a hot dense plasma}

\author{T.C.Luu}
\email[]{tluu@lanl.gov}
\affiliation{Theoretical Division, Los Alamos National Laboratory, MS-227, Los Alamos, New
  Mexico 87545, USA}

\author{J.L. Friar}
\affiliation{Theoretical Division, Los Alamos National Laboratory, MS-227, Los Alamos, New
  Mexico 87545, USA}

\author{A.C. Hayes}
\affiliation{Theoretical Division, Los Alamos National Laboratory, MS-227, Los Alamos, New
  Mexico 87545, USA}

\date{\today}

\begin{abstract}
  In high neutron flux environments where isomers can be strongly
  populated by nucleonic reactions, isotope abundances from reaction
  network chains can be affected by the population of nuclear isomers.
  At high temperatures and densities there is the additional
  possibility of populating these isomers electromagnetically. Here we
  examine the rates for electromagnetic excitation of the isotopes of
  several isomers of interest both in astrophysics and applied physics
  (e.g. $^{235}$U, $^{193}$Ir, and $^{87,88}$Y).  We consider six
  possible electromagnetic processes, namely, photo-absorption,
  inverse internal conversion, inelastic electron scattering, Coulomb
  excitation, $(\gamma,\gamma')$ and $(e,e'\gamma)$ reactions. We find
  that for plasma temperatures $kT\sim 1-10$ keV the electromagnetic
  reactions rates are negligible.  Thus, we conclude that reaction
  network calculations do not need to include for the possibility of
  electromagnetically exciting nuclear isomers. This is true in both
  stellar and terrestrial thermonuclear explosions, as well as in
  plasma conditions expected at the National Ignition Facility.
\end{abstract}
\maketitle

\newpage 

\section{Introduction}
\begin{figure}
  \epsfig{file=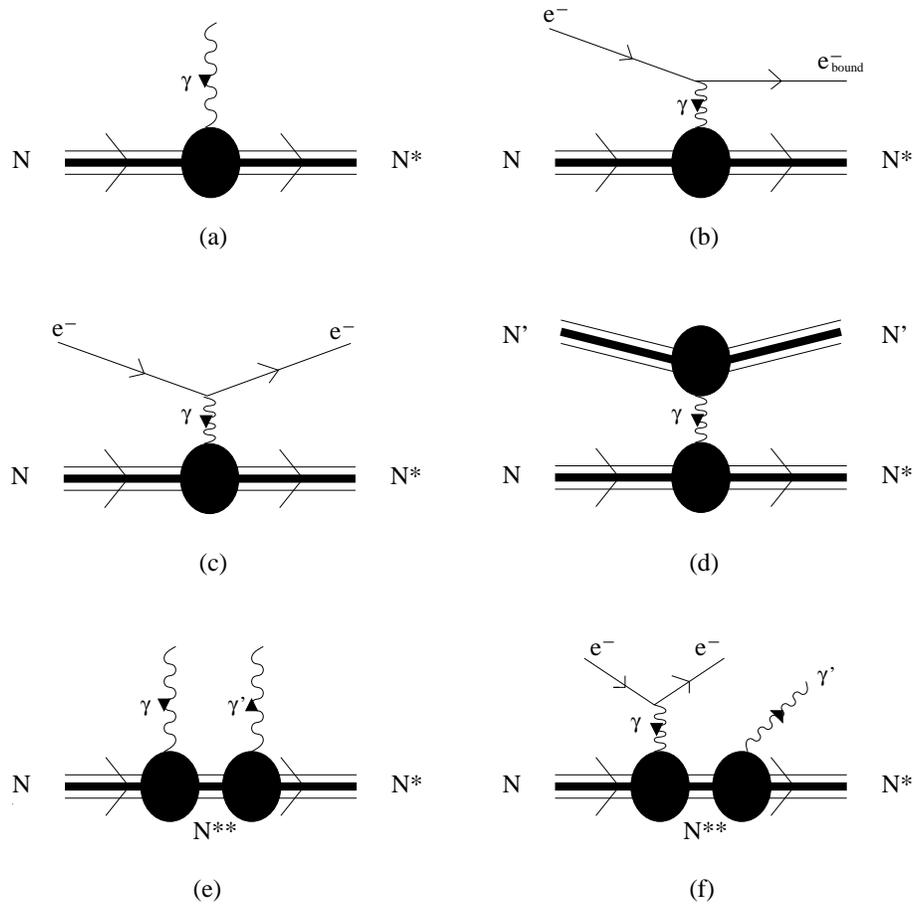,width=12cm}
  \caption{Feynman diagrams representing the electromagnetic processes
  considered in this paper.  Plot (a) represents photo-excitation.
  (b) represents inverse internal conversion, where the final electron
  is bound.  (c) shows electron scattering, where the final electron
  belongs to the continuum.  (d) shows ion-excitation, where N' is
  another ion.  (e) shows ($\gamma,\gamma'$) excitation.  Finally, (f)
  shows ($e,e'\gamma$) excitation.\label{fig1}}
\end{figure}

In a hot dense plasma short-lived excited states of the nucleus with
excitation energies on the order of the temperature of the plasma
reach thermal equilibrium with the nuclear ground state and their
relative population is determined by a Boltzmann
distribution. Nucleosynthesis calculations usually try to take these
states into account, though little is known about their nucleonic
reaction properties (see, for example, Ref\cite{Fowler1980}).
Recently there has been increased interest in the reaction properties
of a certain set of nuclear excited states: nuclear spin isomers.
These isomers are characterized by their relative long lifetimes due
to their large angular momentum compared to their respective ground
states.  Particular interest has been concentrated on the 77 eV isomer
of $^{235}$U, where studies suggest that this isomer is populated
strongly in the $(n,n')$ reaction as well as in the
$^{234}$U$(n,\gamma)$ reaction, and that the fission and neutron
capture cross sections for this isomer are significantly different
than that of the $^{235}$U ground state\cite{Popeko1973,Mostovoi1984}.
Thus, in high neutron flux environments, where two- or multi-neutron
reactions are highly probable, reactions on this isomer can affect the
reaction network chain involving the Uranium isotopes.  The question
we wish to address here is whether nuclear isomers can be populated by
electromagnetic interactions as well.  In a hot dense plasma
environment there are several electromagnetic mechanisms to be
considered at leading and sub-leading order in the fine structure
constant $\alpha$, namely: photo-absorption, inverse internal
conversion, inelastic electron scattering, Coulomb excitation,
$(\gamma,\gamma')$, and $(e,e'\gamma)$\footnote{An omitted reaction at
leading order in $\alpha$ is \emph{inverse} internal pair-production.
The threshold for this reaction occurs at $E>1$ MeV (twice the
electron mass), where $E$ is the excitation energy of the isomer.
However, for the isomers considered here, this condition is not met.
See Ref.\cite{Veres2002} for descriptions of higher order processes.}.
Figure~\ref{fig1} shows the Feynman diagrams for these processes.

The electromagnetic excitation rates for all of these processes
depend, of course, on the temperature $kT$ of the plasma, the
excitation energy $E_m$ of the isomer, and the lifetime $\tau_{1/2}$
of the isomer.  To simplify calculations, we assume that nuclei are
completely stripped of their electrons in the plasma.  Where possible,
we give estimates to the errors of this assumption.  We present
expressions for these electromagnetic processes and examine the rates
for certain nuclear isomers. Of particular interest for applied
physics are the low-lying isomers of $^{235}$U, $^{193}$Ir,
$^{87,88}$Y.  Their properties, as well as those of other isomers
considered in this paper, are listed in Table~\ref{tab:isomers}.  The
half-lives of these isomers range from 0.3 msec to years and their
excitation energies from $\sim 77$ eV - $2.5$ MeV.

\section{Electromagnetic Excitation Rates\label{rates}}

The total reaction width can be expressed as the sum of its
constituent widths:
\begin{equation}\label{eqn:Gamma_tot}
\Gamma_{total}=\Gamma_{\gamma}+\Gamma_{i.c.}+\Gamma_{i.e.s.}\ldots
\end{equation}
Traditionally, this width is expressed
as\cite{Blatt&Weisskopf1979,deShalit&Feshbach1974}
\begin{eqnarray}\label{eqn:width_sum}
  \Gamma_{total}&=&\Gamma_{\gamma}+\alpha_{i.c.}
  \Gamma_{\gamma}+\alpha_{i.e.s.}
  \Gamma_{\gamma}+\ldots \nonumber\\
  &=&\Gamma_{\gamma}(1+\alpha_{i.c.}+\alpha_{i.e.s.}+\ldots).
\end{eqnarray}
Here $i.c.$ refers to internal conversion, $i.e.s.$ refers to
inelastic electron scattering, and so forth.  As can be seen from
Eq.~(\ref{eqn:width_sum}), calculation of a particular channel width
$\Gamma_{i}$ involves calculating both $\Gamma_{\gamma}$ and
$\alpha_{i}$,where $\alpha_{\gamma}=1$ for the case of
photo-absorption.  The reaction rate per nucleon for a particular
channel $i$ is given by
\begin{equation}\label{eqn:rate}
  R_{i}=\int \sigma_{i}(E) d(\Phi(E)) 
\end{equation}
where $\Phi(E)$ is the flux of incoming particles responsible for that
particular reaction and $\sigma_{i}(E)$ is the cross section.  Due to
the relatively long lifetimes of these isomers (\emph{i.e.}
$\Gamma_{total}\ll E_m$), the cross section is well approximated by
\begin{equation}\label{eqn:BW_approx}
  \sigma_{i}(E) \stackrel{\Gamma_{tot}\rightarrow 0}{\approx} \frac{2
  \pi^2}{k^2} \Gamma_{i} \delta(E-E_m),
\end{equation}
where $E_m$ is the excitation energy of the isomer and $k$ is the
wavenumber for the incoming particle.

\subsection{Photo-absorption}
We assume that in the system of interest the plasma has reached
thermal equilibrium so that the photon flux is described by a Plankian
distribution,
\begin{equation}\label{eqn:plankian}
  \begin{split}
    d(\Phi_{\gamma}(E))=&\ \frac{c}{\pi^2(\hbar
    c)^3}\frac{E^2}{e^{E/kT}-1}dE \\ \equiv&\ c N(E) dE,
  \end{split}
\end{equation}
where $N(E)$ is the photon number density in the energy window between
$E$ and $E+dE$.  The photo-absorption rate per unit atom is then
\begin{align}
  R_\gamma =&\ c \int_{0}^{\infty}N(E)\sigma_\gamma(E)dE
  \label{eqn:R_gamma}\\ \sim&\
  \frac{2}{e^{E_m/kT}-1}\frac{\Gamma_\gamma}{\hbar}\label{eqn:R_gamma
  gamma}.
\end{align}
Here $\sigma_\gamma(E)$ is the photo-absorption cross section.
Equation~\ref{eqn:R_gamma gamma} follows from substituting
Eq.~\ref{eqn:BW_approx} into Eq.~\ref{eqn:R_gamma}.  In the
long-wavelength approximation (which is valid for the plasma
conditions considered here) the width $\Gamma_\gamma$ can be expressed
in terms of reduced multipolarity transition probabilities
$B(\pi\lambda)$,
\begin{equation}\label{eqn:widths}
    \Gamma_\gamma(\pi\lambda: J_i\rightarrow J_f)=\ \frac{8\pi
    (\lambda+1)}{\lambda [(2\lambda+1)!!]^2} \left(\frac{E_m}{\hbar
    c}\right)^{2\lambda+1}B(\pi\lambda: J_i\rightarrow J_f)
\end{equation} 
In general, calculating $B(\pi\lambda)$ is very difficult because it
depends on the initial- and final-state nuclear wavefunctions, which
are not typically known.  Naive estimates can be made if one invokes
the extreme single-particle approximation, giving the well-documented
\emph{Weisskopf} expressions\cite{deShalit&Feshbach1974},
\begin{equation}\label{eqn:weisskopf estimates}
  \begin{split}
    B(E\lambda: J_i\rightarrow J_f)=&\
    \frac{1}{4\pi}\left(\frac{3}{3+\lambda}\right)^2 R^{2\lambda}\
    e^2\ \mbox{fm}^{2\lambda}\\ B(M\lambda: J_i\rightarrow J_f)=&\
    \frac{10}{\pi}\left(\frac{3}{3+\lambda}\right)^2 R^{2\lambda-2}\
    \mu_N^2 \mbox{fm}^{2\lambda-2},
  \end{split}
\end{equation}
where $R\sim 1.2\ A^{1/3}$ and $\mu_N$ is the Bohr magneton.  Near
closed-shell nuclei, Weisskopf estimates give reasonable agreement
with experimental values.  However, for nuclei that exhibit strong
collective degrees of freedom, the Weisskopf estimates are too small,
sometimes by a few orders of magnitude\cite{Wong1998}.  

Since the half-life of the isomer of $^{235}$U is about 26 minutes,
temperatures of $\sim 150$ keV would be required for the reaction rate
per unit atom to reach a value of one-per-second. Thus,
photo-absorption does not provide a significant mechanism for 
populating the 77 eV isomer of $^{235}$U.  As can be seen from
Table~\ref{tab:rates}, the photo-absorption rates for all the isomers
considered here are negligible.  Even with more accurate Weisskopf
values, the photo-absorption rates of these isomers at these
temperatures would still be very small.

\subsection{Inverse Internal Conversion}

In inverse internal conversion an electron from the plasma is captured
into a bound atomic orbital, thereby exciting the nucleus. The capture
energy of the electron must match the excitation energy of the nuclear
isomer. The reaction rate is determined by the overlap of the
continuum electron wavefunction with the bound atomic electron
wavefunction, weighted by the nuclear transition matrix element.
Since we assume that nuclei are stripped of all their electrons, all
atomic shells (\emph{i.e.} $K$, $L_I$, $L_{II}$, $M_I$, etc) can
participate in the internal conversion process, as long as the
kinematics of the process are satisfied.  However, since the
conversion happens near the nucleus, s-waves states typically dominate
the transition.  In most cases the electron can capture into the
$1s_{1/2}$ $K$-shell.  For isomers with very low excitation energies the
electron can be captured only into a higher orbit.  For example, the
77 eV isomer of $^{235}$U can capture electrons only into very
high $(n\ge 39)$ outer atomic orbits, whereas both $^{242}$Am and
$^{193}$Ir can capture electrons only into orbits with $n\ge 2$.  

To obtain naive estimates of $\alpha_{i.c.}$, we assume relativistic
hydrogenic wavefunctions for the bound electron and treat the nucleus
as a point particle.  The internal conversion coefficients for the
innermost atomic orbits are thus given by\cite{Rose1958}
\begin{equation}\label{eqn:alphaiic}
  \begin{split}
    \alpha_{i.c.}(E\lambda;\ \kappa_o=-1)= &\ \frac{2 \pi
      \alpha\omega_\gamma}{\lambda(\lambda+1)}
    \sum_{\kappa}(2j+1) 
    \left(
      \begin{array}{ccc}
        \frac{1}{2} & j & \lambda\\ 
        \frac{1}{2} & -\frac{1}{2} & 0
      \end{array}
    \right)^2
    |R_{-1,\kappa}(E\lambda)|^2\\
    \alpha_{i.c.}(M\lambda;\ \kappa_o=-1)= &\ \frac{2 \pi
    \alpha\omega_\gamma}{\lambda(\lambda+1)}
  \sum_{\kappa}(2j+1)(\kappa-1)^2
  \left(
    \begin{array}{ccc}
      \frac{1}{2} & j & \lambda\\ 
      \frac{1}{2} & -\frac{1}{2} & 0
    \end{array}
  \right)^2
  |R_{-1,\kappa}(M\lambda)|^2,
  \end{split}
\end{equation}
where $j=|\kappa|-1/2$ and $\omega_\gamma$ is the photon wave number.
For $M\lambda$ transitions, $\kappa$ can take on values of $-\lambda$
or $\lambda+1$, while for $E\lambda$ transitions, $\kappa=\lambda$ or
$-(\lambda+1)$.  The functions $R_{\kappa_o,\kappa}$ are given by
\begin{equation}\label{eqn:Rfunctions}
  \begin{split}
    R_{\kappa_o,\kappa}(E\lambda)=&(\kappa_o-\kappa)(R_3+R_4)+
    \lambda(R_4-R_3)+\lambda(R_1+R_2)\\ 
    R_{\kappa_o,\kappa}(M\lambda)=&R_5+R_6,
    \end{split}
\end{equation}
where
\begin{equation}\label{eqn:R_ns}
  \begin{split}
    R_1=&\int_0^{\infty}dr\ u_{\kappa}(p_er)U_{\kappa_o}(\lambda_e r)
    h_\lambda^{(1)}(\omega_{\gamma} r)\\
    R_2=&\int_0^{\infty}dr\ v_{\kappa}(p_er)V_{\kappa_o}(\lambda_e r)
    h_\lambda^{(1)}(\omega_{\gamma} r)\\
    R_3=&\int_0^{\infty}dr\ v_{\kappa}(p_er)U_{\kappa_o}(\lambda_e r)
    h_{\lambda-1}^{(1)}(\omega_{\gamma} r)\\
    R_4=&\int_0^{\infty}dr\ u_{\kappa}(p_er)V_{\kappa_o}(\lambda_e r)
    h_{\lambda-1}^{(1)}(\omega_{\gamma} r)\\\
    R_5=&\int_0^{\infty}dr\ v_{\kappa}(p_er)U_{\kappa_o}(\lambda_e r)
    h_\lambda^{(1)}(\omega_{\gamma} r)\\
    R_6=&\int_0^{\infty}dr\ u_{\kappa}(p_er)V_{\kappa_o}(\lambda_e r)
    h_\lambda^{(1)}(\omega_{\gamma} r).
  \end{split}
\end{equation}
Here $U$ and $V$ correspond to the large and small components of the
relativistic radial wavefunction for the bound electron in a Coulombic
field, while $u$ and $v$ are the corresponding analogues for the
Coulomb-distorted continuum wavefunction, and $h_\lambda$ is the
spherical Hankel function of order $\lambda$.  Closed analytic forms
can be derived for these integrals and are given, for example, in
Refs.\cite{Hamilton1975,carroll1965}.

Internal conversion coefficients for the innermost s-wave reactions
are given in Table~\ref{tab:IICcoefficients}.  It is interesting to
note that though internal conversion is suppressed by an order
$\alpha$ compared to photo-absorption, there can be instances where
there are enhancements to the reaction rate.  This is clearly evident
in $^{193}$Ir, $^{242}$Am, and especially $^{235}$U.  This behavior,
though not apparent from Eq.~\ref{eqn:alphaiic}, comes from the
observation that the coefficients scale roughly as\cite{Alder1956}
\begin{displaymath}
  \alpha_{i.c.}\sim \left\{\begin{array}{ll}
  \left(\frac{1}{\omega_{\gamma}}\right)^{\lambda+5/2} & \mbox{for
  $E\lambda$}\\ \left(\frac{1}{\omega_{\gamma}}\right)^{\lambda+3/2} &
  \mbox{for $M\lambda$}
    \end{array}\right.
\end{displaymath}
Since the excitation energies of these isomers are relatively small,
this enhancement is not surprising.

In the systems of interest (\emph{e.g.} ignited inertial confinement
fusion (ICF) capsules, nuclear explosions, and stellar environments),
the temperatures attainable by the plasma are typically not high
enough for nuclei to be completely stripped of their electrons, so
that not all atomic shells are accessible for inverse internal
conversion.  Screening effects arising from the partial stripping of
these nuclei affect the isomer excitation reaction rates by as much as
50\% or more\footnote{If the nucleus is not completely stripped,
reaction methods can also proceed via
NEET\cite{Morita1973,Claverie:2004zv} (Nuclear Excitation by
Electronic Transition). However, this method is even higher order in
$\alpha$, and for most of the isomers considered here, their
excitation energies prohibit this excitation channel.}.
Self-consistent Dirac-Hartree-Fock calculations\cite{Band1993} have
been performed to obtain internal conversion coefficients on some of
the isomers considered here.  These calculations include the full
effects of screening since the nuclei are not considered stripped of
their electrons.  Finite-size effects of the nucleus are considered as
well. Hence these calculations use more realistic electronic
wavefunctions than the simplistic hydrogenic wavefunctions considered
here. We list these results in Table~\ref{tab:IICcoefficients} for
comparison. In general, calculations can differ by an order of
magnitude (or more), as is evident for the case of $^{89m}$Y.  Yet the
overall trend in relative sizes is consistent between the two methods.

The density of electrons is given by
\begin{equation}\label{eqn:elect_density}
  n_e=\frac{m_e^{3/2}}{\sqrt{2}\pi^2\hbar^3}\int dE
  \frac{\sqrt{E}}{e^{(E-\mu)/kT}+1}
\end{equation}
Since there is no simple closed expression for the integral above, the
chemical potential $\mu$ and density must be numerically solved
self-consistently.  The flux $\Phi_e$ of electrons is simply
\begin{equation}\label{eqn:elect_flux}
  \begin{split}
    \Phi_e=&\ \frac{1}{(2\pi\hbar)^3 m_e}\int_{v_z>0} dp^3\frac{p\ 
    cos(\theta)}{e^{(E(p)-\mu)/kT}+1}\\
         =&\ \frac{m_e}{4\pi^2\hbar^3}\int dE
    \frac{E}{e^{(E-\mu)/kT}+1}
  \end {split}
\end{equation}
Hence the differential flux is
\begin{equation}\label{eqn:elect_diff_flux}
  d(\Phi_e)=\ \frac{m_e}{4\pi^2\hbar^3}
    \frac{E}{e^{(E-\mu)/kT}+1} dE
\end{equation}
Using Eqs.~\ref{eqn:rate}-\ref{eqn:BW_approx} gives the rate for
isomeric production via internal conversion as
\begin{equation}\label{eqn:iicrate}
  \frac{\alpha_{i.c.}}{4}\frac{1}{E^{(E_m-\mu)/kT}+1}
  \frac{\Gamma_\gamma}{\hbar}.
\end{equation}

To obtain estimates, we take an average electron density of
$n_e\sim100 N_A\ \mbox{cm}^{-3}$, where $N_A$ is Avogadro's number.
This corresponds to a chemical potential of $\mu=-.00275$ keV
($-.0622$ keV) at $kT=1$ keV ($10$ keV).  Rates are listed in
Table~\ref{tab:rates}. Even with the more accurate internal conversion
coefficients of Ref.\cite{Band1993}, rates would still be negligible.

\subsection{Inelastic electron scattering}

Formally, there is little difference between the theory of inelastic
electron scattering and internal conversion.  Physically, the main
difference is that the final state electron still belongs to the
continuum, as opposed to the bound-state orbital characteristic of
inverse internal conversion.  If the point-nucleus approximation is
made, calculations of `inelastic scattering coefficients' are
identical to the case with internal conversion coefficients--one just
replaces the bound state wavefunctions with Coulomb distorted
continuum wavefunctions, \emph{i.e.}
\begin{equation}
  \begin{split}
    R_1=&\int_0^{\infty}dr\ u_{\kappa}(pr)u_{\kappa_o}(p_o r)
    h_\lambda^{(1)}(\omega_{\gamma} r)\\
    R_2=&\int_0^{\infty}dr\ v_{\kappa}(pr)v_{\kappa_o}(p_o r)
    h_\lambda^{(1)}(\omega_{\gamma} r)\\
    R_3=&\int_0^{\infty}dr\ v_{\kappa}(pr)u_{\kappa_o}(p_o r)
    h_{\lambda-1}^{(1)}(\omega_{\gamma} r)\\
    R_4=&\int_0^{\infty}dr\ u_{\kappa}(pr)v_{\kappa_o}(p_o r)
    h_{\lambda-1}^{(1)}(\omega_{\gamma} r)\\
    R_5=&\int_0^{\infty}dr\ v_{\kappa}(pr)u_{\kappa_o}(p_o r)
    h_\lambda^{(1)}(\omega_{\gamma} r)\\
    R_6=&\int_0^{\infty}dr\ u_{\kappa}(pr)v_{\kappa_o}(p_o r)
    h_\lambda^{(1)}(\omega_{\gamma} r).
  \end{split}
\end{equation}
The above integrals can be computed numerically and the coefficients
calculated using Eq.~\ref{eqn:alphaiic}.  Note that since these
calculations deal with initial- and final-state \emph{continuum}
wavefunctions, coefficients no longer have any dependence on principal
quantum numbers $n$.  Nor do they take on discrete values since they
are now functions of the electron's initial energy $E$, which is
continuous.  Since we approximate the reaction rate as being strongly
peaked at the isomer's excitation energy (\emph{i.e.}
Eq.~\ref{eqn:BW_approx}), the reaction rate is simply
\begin{equation}\label{eqn:iesrate}
  \frac{\alpha_{i.e.s.}(E_m)}{4}\frac{1}{E^{(E_m-\mu)/kT}+1}
  \frac{\Gamma_\gamma}{\hbar}.
\end{equation}
Lastly, we assume the outgoing electron to be in an s-wave state.
Table~\ref{tab:IICcoefficients} tabulates the inelastic scattering
coefficients at each isomer's excitation energy
$E_m$. Table~\ref{tab:rates} shows rates for this reaction channel
using the same conditions for the electron flux as in the previous
section.

\subsection{Coulomb excitation via heavy-ion scattering\label{CE}}

Isomeric excitation could also happen via heavy-ion Coulomb
excitation.  In this case, the kinetic energy of the ions is small
compared to its rest mass.  Hence the ions can be treated
non-relativistically.  Reaction rates for heavy-ion Coulomb
excitations can be directly calculated using similar methods discussed
in the previous sections.  However, it is simpler (and perhaps more
enlightening) to consider the `adiabaticity'
parameter\cite{Wong1998,Brussaard1965}
\begin{equation}\label{eqn:adiab}
  \xi=\eta\frac{E_{m}}{E},
\end{equation}
where $E$ is the incoming ion energy and $\eta$ is the classical
Sommerfeld number given by
\begin{equation}\label{sommer}
    \eta=\alpha\ Z_1\ Z_2\ \frac{c}{v}.
\end{equation}
Here $Z_1$ and $Z_2$ are the charges of the two scattering ions.  The
parameter $\xi$ is a measure of the ratio between the collision time
and the nuclear period.  The larger the value of $\xi$, the more
adiabatic the process becomes.  Under conditions where the reaction is
highly adiabatic, excitation of the nucleus is unlikely due to the
fact that the nucleus has time to equilibrate under the influence of
the external Coulomb field.  Assuming the ions are in thermal
equilibrium, the energy $E$ can be approximated by $\frac{3}{2}kT$.
Considering like-ion scattering, \emph{i.e.} $Z_1=Z_2$, the range of
temperatures considered in this paper give
\begin{displaymath}
  10^3 <\  \xi\  <10^7.
\end{displaymath} 
The cross section for heavy-ion scattering scales roughly as $\sigma
\sim e^{-2\pi\xi}$\cite{Brussaard1965}, which in turn gives a similar
scaling for the reaction rate.  This rate is extremely small for large
$\xi$. It is interesting to note that the exponential suppression is
primarily a property of the \emph{repulsive} Coulomb
field\cite{Brussaard1965}. The Coulomb barrier of these ions is too
strong at these plasma temperatures for any isomeric excitation.  For
an attractive field with $Z_1=1$ (as in inelastic electron
scattering), this exponential behavior is not as pronounced.

\subsection{N($\gamma$,$\gamma'$)N* and N($e$,$e'\gamma$)N*\label{Ngamma}}
Another possible excitation process is through excitation to some high
lying state, followed by gamma decay to the isomer.  Excitation can be
done via all the processes mentioned above, and in particular through
($\gamma$,$\gamma'$) and ($e$,$e'\gamma$) reactions.  If one assumes
that the rate of decay to the isomer is fast compared to the
excitation process, then the rate of producing the isomer is
approximately equal to the rate of excitation to the high-lying state.
At first glance, such a process might seem plausible, since many of
these higher-lying states are connected to the ground state by
transitions of low multipolarity ($\lambda<$3,4).  Such probability
transitions may be larger by an order of magnitude or so compared to
the isomeric transitions.  However, the fluence of incoming particles
with the necessary energy to allow excitation is exponentially
suppressed.  Hence the gain from going to a lower multipolarity
transition is quickly lost due to the fact that these high-lying
states have large excitation energies compared to the isomers.

For example, consider the case of $^{89}_{39}Y$.  The isomer state is
connected via M1 transition to the $7/2^+$ state at energy 2529.87
keV.  The transition probability between these two states is large
($\tau_{1/2}$=.08 ps).  The $7/2^+$ state can be connected to the
ground state via an M3 transition.  But since the energy of this high
lying state is large compared to the isomer, the rate of populating
this state is much smaller than for the rate of directly populating
the isomer.  This suppression is roughly
\[
\frac{R_{7/2^+}}{R_{isomer}}\approx\frac{e^{-2530/kT}}{e^{-909/kT}}
=e^{-1621/kT}.
\]
Here $kT$ is in units of keV and $R_{isomer}$ is the rate for
producing the isomer directly from the ground state (via M4
transition).  For a plasma temperature of 1 keV/10 keV, this
suppression factor is vanishingly small.  Similar suppression factors
are found for the other isomers considered in this paper.  Hence we
conclude that population of isomers via $(\gamma,\gamma')$ or
$(e,e'\gamma)$ excitation is extremely unlikely (as opposed to the
case of neutron-induced excitations, for example).

\section{Conclusion\label{conclusion}}
Nuclear isomers have played an increasingly important role in both
astrophysics and applied physics phenemona.  Typically these isomers
are assumed to be populated solely by strong nuclear reactions.
However, if the temperatures of the environment in which the isomers
exist are high enough, there is the possibility that they be populated
electromagnetically as well.  In this paper we addressed this latter
issue for typical plasma temperatures found in mundane laboratories
and astrophysical environments (\emph{i.e.}  kT$\sim$ 1 to 10 KeV).

In Sect.~\ref{rates} we gave expressions for the reaction rates of the
following electromagnetic processes: photo-absorption, inverse
internal conversion, inelastic electron scattering, and Coulomb
excitation.  In Sect.~\ref{Ngamma} we argued that rates for
$(\gamma,\gamma')$ and $(e,e'\gamma)$ excitation were negligible.  We
made the assumption that the nucleus was a point particle and was
completely stripped of its electrons due to the plasma
environment. Furthermore, we only considered s-wave scattering since
this is usually the dominant channel.  We found that all these
processes contributed negligibly to the overall reaction rates for
populating the nuclear isomers listed in Table~\ref{tab:isomers}.
Table~\ref{tab:rates} summarizes our claims.

These small rates can also be understood from simple dimensional
analysis. As an example, consider photo-absorption on the ground state
of $^{235}$U.  Using Eqns.~\ref{eqn:R_gamma gamma}-\ref{eqn:weisskopf
estimates} as guides, naive dimensional analysis gives the reaction rate
for producing Uranium isomers as 
\begin{equation}\label{eqn:dim_anal}
R_\gamma\sim\frac{kT}{\hbar\omega_\gamma}\ \alpha\ Q^{2 l}\
\frac{\omega_\gamma}{\hbar},
\end{equation}
where in this case $l$=3 and $Q=\frac{\omega_\gamma R}{c}\sim
2.4\times 10^{-6}$.  The dimensionless parameter $Q$ sets the scale
of the transition.  Similar expressions for internal conversion and
electron scattering can be determined from dimensional analysis as
well, though in these cases it is the electron's momentum that
determines the size of $Q$ and not the photon's momentum.  In all of
these cases, $Q$ is very small.  An exception would be Coulomb
excitation.  However, as already mentioned in Sect.~\ref{CE}, the
Coulomb barrier is prohibitive at the temperatures considered here.

Clearly the assumption that the nucleus is completely stripped of
electrons is not valid.  Furthermore, the expressions given in
Sect.~\ref{rates} were made under the assumption that the fluence of
incoming particles is in thermal equilibrium.  This is most likely not
true in applied physics applications.  However, due to the extremely
small rates listed in Table~\ref{tab:rates}, more sophisticated
calculations that take these issues into consideration are not
warranted.  Hence the assumption that nuclear isomers are produced
entirely via strong nuclear reactions for the range of temperatures
considered in this paper is valid.

\newpage


\newpage

\begin{table}[h]
\caption{Nuclear isomers and their properties.\label{tab:isomers}}
\begin{tabular}{|c|c|c|c|c|c|}
\hline Nucleus & ground state (J$^\pi$) & multipole & isomer (J$^\pi$)
& excitation energy $E_m$ & $\tau_{1/2}$\\ \hline $^{87}$Y& $1/2^-$ &
M4 & $9/2^+$ & 380.7 keV & 13.37 hr\\ $^{88}$Y & $4^-$ & E3 & $1^+$ &
392.9 keV & 0.3 ms\\ $^{88}$Y & $4^-$ & M4 & $8^+$ & 674.6 keV & 13.9
ms \\ $^{89}$Y & $1/2^-$ & M4 & $9/2^+$ & 909 keV & 16.06 s\\
$^{178}$Hf & $0^+$ & M8 & $8^-$ & 1.174 MeV & 4 s\\ $^{178}$Hf & $0^+$
& E16 & $16^+$ & 2.446 MeV & 31 yr\\ $^{193}$Ir&$3/2^+$ & M4 &
$11/2^-$& 80.2 keV & 10.53 d\\ $^{235}$U &$7/2^-$ & E3 & $1/2^+$ & 77
eV & 26 min \\ $^{242}$Am &$1^-$ & E4 & $5^-$ & 48.63 keV & 141 yr \\
\hline
\end{tabular}
\end{table}

\newpage

\begin{table}[h]
\caption{Internal conversion and inelastic scattering coefficients.
The second column gives the orbital in which the electrons are
captured during inverse internal reactions.  The third column lists
the conversion coefficients calculated from relativistic hydrogenic
wavefunctions, (\emph{i.e.} Eqs.~\ref{eqn:alphaiic}-\ref{eqn:R_ns}).
The fourth column gives conversion coefficients derived from
Dirac-Hartree-Fock calculations.  The last column gives scattering
coefficients for s-wave electrons calculated at the isomer's
excitation energy $E_m$.
\label{tab:IICcoefficients}}
\begin{tabular}{|c|c|c|c|c|}
\hline Isomer & final electron state & $\alpha_{i.c.}$ (r.h.wfs) &
$\alpha_{i.c.}$ (D.H.F.)\footnote{Ref.\cite{Band1993}} &
$\alpha_{i.e.s.}(E_m)$ \\ \hline 
$^{87m}$Y & $1$s$_{1/2}$ & .227 & & 3.54 \\ 
$^{88m1}$Y & $1$s$_{1/2}$ & .0245 & & 1.48 \\ 
$^{88m2}$Y & $1$s$_{1/2}$ & .0226 & & .323 \\ 
$^{89m}$Y & $1$s$_{1/2}$ & .0078 & .0819 & .107\\ 
$^{178m1}$Hf & $1$s$_{1/2}$ & .342 & & 3.05\\
$^{178m2}$Hf & $1$s$_{1/2}$ & .109 & & 20.47\\ 
$^{193m}$Ir & $2$s$_{1/2}$ & 3641 & 2610 & 1.9E5\\ 
$^{235m}$U & $39$s$_{1/2}$ & 1.5E17 & 2.62E17\footnote {O$_{VI}$ shell
  result} & 1.9E22\\ 
$^{242m}$Am & $2$s$_{1/2}$ & 3724 & & 4.5E5 \\ \hline
\end{tabular}
\end{table}

\newpage

\begin{table}[h]
\caption{Order of magnitude rates for electromagnetic reactions
involving photoabsorption (P.A.), inverse internal conversion
(I.I.C.), inelastic electron scattering (I.E.S.), and Coulomb
excitation (C.E.).  Rates are in units of inverse seconds, and come in
pairs.  The first number corresponds to the rate at $kT=1$ keV, while
the second to $kT=10$ keV.  We do not list rates for
($\gamma,\gamma'$) and ($e,e'\gamma$) since these are vanishingly
small as well (see discussion in Sect.~\ref{Ngamma})
. \label{tab:rates}}
\begin{tabular}{|c|c|c|c|c|}
\hline Isomer & P.A.  & I.I.C. & I.E.S.  & C.E. \\
\hline 
$^{87m}$Y& $10^{-171}|\ 10^{-22}$ & $10^{-173}|\ 10^{-26}$& $10^{-172}
|\ 10^{-25}$ & $\sim e^{-10^7}|\ e^{-10^6}$\\ 

$^{88m1}$Y & $10^{-168}|\ 10^{-14}$& $10^{-172}|\ 10^{-20}$&
$10^{-170}|\ 10^{-18}$ & $\sim e^{-10^7}|\ e^{-10^5}$\\ 

$^{88m2}$Y & $10^{-296}|\ 10^{-32}$& $10^{-300}|\ 10^{-38}$&
$10^{-299}|\ 10^{-37}$ & $\sim e^{-10^7}|\ e^{-10^6}$ \\ 

$^{89m}$Y & $10^{-397}|\ 10^{-41}$ & $10^{-401}|\ 10^{-47}$&
$10^{-399}|\ 10^{-46}$ & $\sim e^{-10^7}|\ e^{-10^6}$ \\ 

$^{178m1}$Hf & $10^{-505}|\ 10^{-70}$& $10^{-522}|\ 10^{-75}$&
$10^{-521}|\ 10^{-74}$ & $\sim e^{-10^8}|\ e^{-10^7}$ \\ 

$^{178m2}$Hf & $10^{-1116}|\ 10^{-160}$& $10^{-1119}|\ 10^{-165}$&
$10^{-1118}|\ 10^{-164}$& $\sim e^{-10^7}|\ e^{-10^6}$ \\ 

$^{193m}$Ir & $10^{-46}|\ 10^{-14}$ & $10^{-44}|\ 10^{-14}$ &
$10^{-42}|\ 10^{-12}$ & $\sim e^{-10^7}|\ e^{-10^6}$ \\ 

$^{235m}$U & $10^{-21}|\ 10^{-20}$& $10^{-9}|\ 10^{-7}$ & $10^{-4}|\
10^{-2}$  & $\sim e^{-10^4}|\ e^{-10^3}$ \\ 

$^{242m}$Am & $10^{-31}|\ 10^{-12}$& $10^{-30}|\ 10^{-12}$ &
$10^{-28}|\ 10^{-10}$ & $\sim e^{-10^7}|\ e^{-10^6}$ \\ 
\hline
\end{tabular}
\end{table}

\newpage

\end{document}